\documentstyle[aps,prd,preprint]{revtex}
\newfont{\euler}{eufm10 scaled \magstep1}

\begin{document}

\draft
\tightenlines

\preprint{November 1994 \hspace{10cm}
\parbox[t]{10em}{ KUL-TF-94/38 \newline
         CERN-TH.7477/94 \newline
         hep-th/9411221} }
\title{On the Lagrangian Realization of Non-Critical ${\cal W}$-Strings}
\author{Alex Deckmyn\cite{ADRS},
             Ruud Siebelink\cite{ADRS},
             Walter Troost\cite{WT}}
\address{Instituut voor Theoretische Fysica,
KU Leuven,
Celestijnenlaan 200D, B-3001 Leuven, Belgium}
\author{and Alexander Sevrin\cite{AS}}
\address{CERN}
\date{\today}
\maketitle
\begin{abstract}
A large class of non-critical string theories with extended worldsheet
gauge symmetry are described by two coupled, gauged Wess-Zumino-Witten
Models. We give a detailed analysis of the gauge invariant action and
in particular the gauge fixing procedure and the resulting BRST
symmetries.
The results are applied to the example of ${\cal W}_3$ strings.

\end{abstract}

\section{Introduction}

Whereas the simplest models in string theory are based on the
Virasoro algebra or supersymmetric extensions thereof, a lot of
interest has been generated by extensions based on non-linear
symmetry algebras \cite{ZamW3}, called ${\cal W}$ algebras. There
are several lines of investigation for systems having an extended
conformal symmetry. One possibility is to make use of the symmetry
algebras only, trying to gain information about their
representations, and in this way about the possible physical string
models these non-linear algebras correspond to. This is an ambitious
line, but probably still too difficult at the present time. More or
less complete data about representations have till now only been
obtained for some simple finite analogues of these ${\cal W}$
algebras \cite{finiteW}, and for ${\cal W}_3$ \cite{W3rep}. A
different approach has been to realize the Operator Product
Expansions of the ${\cal W}$ algebras in terms of free fields ---
which are easily realized in Fock-space --- and investigate physical
consequences (BRST operators and their cohomology) in these
realizations. In this paper we follow a third line, related to the
previous one, and accord a central role to Lagrangian realizations
of the symmetry algebras in terms of Wess-Zumino-Witten models. This
is done first on a classical level, after which the theories
described by these Lagrangians can be quantized. The transition to
quantum theory is in practice very simple: it amounts to assuming
the validity of affine Lie algebra OPEs for the symmetry currents of
the theory. Moreover, these models are very malleable in that, by
gauging and constraining, they allow the construction of (almost?)
all extended conformal algebras.

One has to distinguish between critical and non-critical models. The
critical models impose a cancellation between the central charges of
the ``matter'' component of the model against the ``ghost''
particles (implying, for example, for the simplest bosonic string a
central charge $c=26$ and for a model based on the ${\cal W}_3$
algebra a value $c=100$). The non-critical models achieve this
cancellation by introducing another sector, the gravitational
sector. This can be understood from the fact that integrating over
matter and ghosts first induces, through a quantum anomaly, an
action for classically non-existing degrees of freedom. For the
simple bosonic string in the conformal gauge this induced action is
the Liouville action, whence it is also called the ``Liouville"
sector. The induced action describes an extension of two dimensional
gravitation theory. The subsequent integration over its degrees of
freedom restores the non-linear symmetry of the theory.

Non-critical ${\cal W}$ string theories were first constructed ``by hand''
\cite{W3BRST}, meaning that
the symmetry currents of both the matter and gravity sectors are
realized in terms of free fields and the BRST operator is then
constructed by trial and error. Though this is quite feasible for the
simplest models, it turns out to be a formidable task for more
complicated models. Obviously, a more systematic apprach is needed.
Recently several possible approaches were discovered.

A most elegant way to solve extended non-critical string theories is
by using the (suspected) equivalence of a large class of them, the
so-called $(1,q)$ models, to topological stringtheories \cite{top}.
Using the matter picture \cite{tokyo,wlas} for these topological
strings, choosing a Landau-Ginzburg type realization of the matter
sector provides a very quick way to investigate several essential
properties, such as the spectrum, of the non-critical string theory.

A related approach takes advantage of the hidden $N{=}2$ structure of
any string theory. The BRST current and the Virasoro anti-ghost
together provide the two supercharges \cite{bea,BLNW,blls}. Adopting
this as the essential structure of any string theory, one then views
the construction of string theories as the study of realizations and
representations of extensions of the $N{=}2$ conformal algebra. This
implies then that one should be able to construct a large class of
non-critical string theories from Hamiltonian reduction. Indeed, many
$N{=}2$ algebras can be constructed by the reduction of WZW models on
supergroups, the reduction being determined by an embedding of
$SU(2|1)$ in a supergroup. By an appropriate choice of the grading ---
which is necessary to determine the reduction completely --- one
obtains a certain free-field realization which can immediately be
viewed as a non-critical string theory. Though this approach looks
very elegant and promising, it has only been established in certain
cases \cite{blls}.

A last approach, again relying on gauged or reduced WZW models, takes
reduced WZW models for both the matter and the gravity sector
separately. Precisely this approach will be studied here.

In this paper we will exploit the versatility of the WZW models.
First, in section \ref{Wmat:DS}, we will analyse a constrained WZW
model, showing how, following the ideas of the Drinfeld-Sokolov
reduction scheme, one can use them to realize ${\cal W}$ algebras.
Our treatment here improves on the ones existing in the literature
in that the auxiliary fields, necessary to save DS gauge invariance
on the Lagrangian level, now arise as a natural part of the
construction, based as it is on that gauge invariance from the
start. This is shown with the help of a recursion method to perform
the transition to the so-called highest weight gauge, in which the
appearance of the ${\cal W}$ algebra is the most manifest. As a
by-product, we also give an efficient recursive method to construct
the gauge invariant polynomials that realize the ${\cal W}$ algebra.
The constructions in this section are relevant for both critical and
non-critical strings. Then, in section \ref{Wmat:transfo}, we
introduce the transformations of the ${\cal W}$ symmetry. We use the
previous construction in section \ref{non-critW} both for the matter
and the gravity sectors. We show how, already at the classical level,
it is only through a cancellation of central charges of the sectors
that the symmetry is achieved. As an application, we give in
subsection \ref{non-critW:BRST} the expression for the classical
BRST charge for the combined matter-ghost-gravity system in the case
of ${\cal W}_3$ that follows from our construction. Our method gives
an expression for this charge that extends to the quantum theory by
a simple renormalization of a single coefficient, without the need
for any additional terms.

In section \ref{BV} we give a more thorough treatment of the gauge
fixing procedure, using the field-antifield formalism of Batalin and
Vilkovisky. First (\ref{BV:DS}) we use this method in the
realization of a single sector to explicitize the fixing of the DS
gauge invariance. This simplifies the derivation of the gauge fixed
action in \cite{Zfac} as it avoids any explicit reference to open
gauge algebras. Then (\ref{BV:W}) we apply the same method to the
additional ${\cal W}$ symmetry that is present if one combines a
matter and a gravity sector. We keep the discussion general, working
out the ${\cal W}_3$ case explicitly at the end. This serves as a
justification of the ghost Lagrangian used in that (relatively
simple) case in section \ref{non-critW:BRST}, and also points the
way to extend the present treatment to arbitrary extensions that can
be obtained from DS reduction.

A more detailed treatment of the results presented in this paper
can be found in \cite{Ruud,Alex}.

\section{The classical action of ${\cal W}$ matter}\label{Wmatter}

\subsection{The Drinfeld-Sokolov procedure revisited}\label{Wmat:DS}
In this first section, we realize a ${\cal W}$ system (matter or
gravity)
by constraining the currents of a WZW model.
We will not review the method of hamiltonian reduction here ---
we only give a cursory description to establish notation --- and
refer the interested reader to
e.g.\ \cite{Hamred,Zfac} for a general introduction and references.
We will supplement the standard treatment with some detailed
recursion formulas to carry out this reduction in practice, since we
need these for later use.

The starting point is the usual WZW action $\kappa{\cal S}^-[g]$
for some Lie (super)group $G$ with generic element $g(z,\bar{z})$.
The ${\cal W}$ algebra is determined by choosing a particular
$sl(2)$ embedding ${\cal S}=\{e_0,e_+,e_-\}$ in the algebra
{\euler g}. The first step is to constrain the current
$J(z)=\frac{\kappa}{2}\partial g \cdot g^{-1}$ to the form
\begin{equation}
J\mapsto \tilde J =
 \frac{\kappa}{2} \partial\tilde g \cdot\tilde g ^{-1} =
 \frac{\kappa}{2} e_- + \frac{\kappa}{2} [\tau, e_-] +
  J^{\geq 0} ,
\label{vormtJ}
\end{equation}
where $J^{\geq 0}$
denotes the positively graded components in the grading induced by
$e_0$, and $\tau$ is a set of auxiliary fields with grading $1/2$ that
are introduced to insure that all constraints are first class
\cite{BerOog,Zfac}. These constraints generate the
Drinfeld-Sokolov (DS) gauge transformations. They can be used to put
the current $\tilde J $ in the {\em highest weight gauge}:
\begin{equation}
\tilde J  = \frac{\kappa}{2}\partial\tilde g \cdot\tilde g ^{-1}
\mapsto \frac{\kappa}{2}\partial e^{\displaystyle\gamma} \tilde g
\cdot\tilde g ^{-1}e^{\displaystyle -\gamma}
\\[2mm] = \frac{\kappa}{2} e_- + W(\tilde J ) ,
\label{DStrans}
\end{equation}
where $W$ contains only highest weight components. These components
are {\em gauge invariant polynomials} of the original components of
$\tilde J $ and their derivatives, and form a classical ${\cal W}$
algebra under Poisson Brackets. We will denote the gauge fixed group
element by $e^{\gamma }\tilde g =w$. The existence and uniqueness of
the algebra element $\gamma $ defining the transition to the highest
weight gauge has been proven long ago \cite{Hamred}, but we present
here an algorithmic procedure to calculate it exactly. For convenience
we first introduce the notations
$E_- \equiv \mbox{ad}(e_-)$, $E_+ \equiv \mbox{ad}(e_+)$
and furthermore we define the ``inverse'' L of $E_-$ \cite{W3allord},
which vanishes on highest weight generators and
$L E_- = 1$ on $\tilde g /\mbox{ker}(E_-)$ .
The highest weight gauge can now be defined by
\begin{equation}
L\left\{ e^{\displaystyle\gamma} \left( e_- +2J^{\geq 0}/\kappa
+[\tau,e_-]  -\partial
\right)  e^{\displaystyle -\gamma} - e_- \right\} = 0 .
\label{vglgam}
\end{equation}
This equation can be solved order by order in $J^{\geq 0}$  and
$\tau$ by writing
$\gamma = \sum_{n\geq 1} \gamma_n$, $ W = \sum_{n\geq 1} W_n$. Up
to first order the equation becomes
\begin{equation}
L\left\{ -E_-\gamma_1 + \partial\gamma_1  +2J^{\geq 0}/\kappa
+[\tau,e_-]
\right\} = 0 , \label{vglgam1}
\end{equation}
and since $\gamma$ is positively graded (and thus $LE_- \gamma_1=
\gamma_1$) the solution is
\begin{equation}
\gamma_1 = \frac{L}{1-L\partial} \left\{ 2J^{\geq 0}/\kappa +
[\tau,e_-]
\right\}\, . \label{gam1}
\end{equation}
At higher order  one may easily construct the recursive algorithm
\begin{eqnarray}
\gamma_n &=& \frac{L}{1-L\partial} {\cal P}_n
 \left\{ e^{\displaystyle \gamma_1+\cdots +\gamma_{n-1}}
   \phantom{\left( 2J^{\geq 0} \right)} \right.
   \nonumber \\[2mm]
&& \phantom{L} \left.
 \left( e_- +2J^{\geq 0}/\kappa +[\tau,e_-]  -\partial \right)
   e^{\displaystyle -\gamma_1-\cdots -\gamma_{n-1}} \right\}
   \nonumber \\[2mm]
2W_n/\kappa &=& \Pi_{\text{hw}} \frac{1}{1-L\partial}{\cal P}_n
 \left\{ e^{\displaystyle \gamma_1+\cdots +\gamma_{n-1}}
   \phantom{\left( 2J^{\geq 0} \right)} \right.
   \label{finalpol} \\
&& \phantom{L} \left.
 \left( e_- +2J^{\geq 0}/\kappa +[\tau,e_-]  -\partial \right)
   e^{\displaystyle -\gamma_1-\cdots -\gamma_{n-1}} \right\}\ ,
   \nonumber
\end{eqnarray}
where ${\cal P}_n$ indicates that we only retain the part of order
$n$. Notice
that the expansion of $\gamma $ and $W$ terminates after a finite
number of
steps, since the expression
\begin{equation}
{\cal P}_n \left\{ e^{\displaystyle \gamma_1+\cdots +\gamma_{n-1}}
  \left( e_- +2J^{\geq 0}/\kappa +[\tau,e_-]  -\partial \right)
  e^{\displaystyle -\gamma_1-\cdots -\gamma_{n-1}} \right\}
\end{equation}
contains only components of grading $(\frac{n}{2}-1)$ or higher.

The action $ {\cal S}^-[w] $
is obviously invariant under DS gauge  transformations,
as it involves only the gauge invariant polynomials $W$.
In addition, the WZW action ${\cal S}^-[g]$ has, from the start,
an invariance under (left) multiplication of $g$ with an arbitrary
{\em holomorphic} group element. The constraints imposed in the DS
reduction also reduce this additional invariance, namely to the
transformations generated by the DS gauge invariant polynomials $W$.
These are called ${\cal W}$ transformations. One may attempt to lift
the restriction to holomorphic parameters by coupling the $W$ to an
extra external field $\mu$. This will be discussed further in the next
section. This same coupling can also be used to great effect to study
the induced ${\cal W}$ gravity theory itself,
see \cite{KPZ,OSSvN,SSvN,W3allord,Zfac}.
We therefore continue with the action
$S = {\cal S}^-[w]+\int\mu\cdot W$.
The recursion relations derived above
can be used to rewrite it as follows, making explicit the dependences
on the auxiliary field and the WZW currents $J$.
Using $w (\tilde J ) = e^{\displaystyle \, \gamma} \tilde g $,
and splitting the WZW action
$\kappa {\cal S}^-[w]$, with help of the Polyakov-Wiegmann identity
\cite{PolWieg},
\begin{equation}
{\cal S}^-[hg] = {\cal S}^-[h] + {\cal S}^-[g]
 -\frac{1}{2\pi x}\int \mbox{str}
 \left\{ h^{-1}\bar\partial h \, \partial g g^{-1} \right\}\, ,
\label{PolWform}
\end{equation}
we obtain
\begin{equation}
\kappa {\cal S}^-[w] = \kappa {\cal S}^-[\tilde g ]
 + \kappa {\cal S}^-[e^{\displaystyle \, \gamma}]
 + \frac{1}{\pi x} \int \mbox{str}
  \left\{ \bar{\partial}
  e^{\displaystyle -\gamma} \cdot e^{\displaystyle \, \gamma}
  \tilde J \right\} \, \, .
\end{equation}
Since $\gamma$ is strictly positively graded, the WZW action
$\kappa {\cal S}^-[e^{\displaystyle \, \gamma}]$
vanishes identically. The
local mixed term simplifies too, and we find that
\begin{eqnarray}
S &=& \kappa {\cal S}^-[\tilde g ]
 + \frac{1}{\pi x} \int \mbox{str}
  \left\{ \frac{\kappa}{2} \bar{\partial} e^{\displaystyle -\gamma}
   \cdot e^{\displaystyle \, \gamma}
   \left( e_- +[\tau, e_-] \right) +
   \mu W(\tilde J ) \right\}
\nonumber \\[2mm]
&=& \kappa {\cal S}^-[\tilde g ]
 + \frac{\kappa}{4\pi x} \int \mbox{str}
  \left\{ [\tau, e_-] \bar{\partial} \tau \right\}
 + \frac{1}{\pi x} \int \mbox{str}
  \left\{ \mu W(\tilde J ) \right\} {}.
\label{actiesplit}
\end{eqnarray}
To derive this last result we inserted the explicit
expressions for $\tilde J $, $\gamma_1$ and $\gamma_2$ that can be
read off from the equations (\ref{vormtJ}) and (\ref{finalpol}).
Higher order terms of $\gamma$ do not contribute to the supertrace. In
eq.(\ref{actiesplit}) the DS gauge invariance is still present, and
will have to be fixed eventually. This can be done in different ways,
which allows one to derive all order expressions for the induced
${\cal W}$-action, see \cite{W3allord,Zfac}. Remark that the kinetic
term for the auxiliary field $\tau$, added ad-hoc in \cite{Zfac} to
preserve gauge invariance, emerges very naturally in the present
formulation, which is based on that gauge invariance from the start.
We will come back to the gauge fixing in section \ref{BV}.

\subsection{$\cal W$ transformations}\label{Wmat:transfo}

In the previous subsection we introduced a constrained WZW action,
where the (DS) gauge invariance could be used to bring the currents
in a highest weight form. Here, we analyze the $\cal W$
transformations themselves.


Infinitesimally the ${\cal W}$ transformations are of the form
$\delta w = X_w w$ where $X_w \in \text{\euler g}$ should be determined
such that the highest weight gauge
(\ref{vormtJ}) is preserved.
This means that the transformation acts on
the highest weight current $W$ only, so we demand%
{}~\footnote{We denote, for any current $j$, the covariant
derivative as
$D[j ]\equiv \partial -\text{ad}(j)$. Later we will also use
$\bar D[A ]\equiv \bar\partial -\text{ad}(A)$.}
that
\begin{equation}
 L \delta \left( 2W/\kappa \right) =
   L \left( D\left[ 2W/\kappa\right] -E_- \right) X_w = 0 .
\label{varww}
\end{equation}
Defining, for any current $j$, the operator $I[j]$ by%

\begin{equation}
I\left[ j\right] \equiv 1 -LD\left[j \right] ,
\label{defI}
\end{equation}
and using the identity $1 - LE_- = \Pi_{\text{lw}}=$ the projection
operator on lowest weight components,
the general solution for $X_w$ can be written as
\begin{equation}
X_w = \frac{1}{I\left[ 2W/\kappa\right]} \eta \qquad\mbox{with}\qquad
\eta\in\mbox{ker}E_- .
\label{Xwres}
\end{equation}
Notice that the inverse operator
$
\frac{1}{I\left[ 2W/\kappa\right]} \equiv \sum_{i\geq0} \left(
LD\left[
2W/\kappa\right] \right)^i
$
is well-defined since each factor
$LD \left[ 2W/\kappa\right]$
increases the $sl(2)$ grading with at least one unit, so that the
sum, when applied to any current, terminates after a finite number
of steps.

Once we have determined the form of the parameter $X_w$, we can
derive the $\eta$ transformation rules for the highest weight
currents. They can be encoded in the matrix equation
\begin{equation}
\delta W = \frac{\kappa}{2} \Pi_{\text{hw}} D\left[ 2W/\kappa\right]
\frac{1}{I\left[ 2W/\kappa\right]} \eta\, .
\label{delWhw}
\end{equation}
These constraint preserving $\eta$ transformations are
nothing but the ${\cal W}$ transformations, which are generated by
the
$W$ currents themselves through Dirac brackets \cite{Hamred}.
These Dirac brackets are equivalent to the Poisson
brackets of the gauge invariant polynomials discussed above, defining
the
classical ${\cal W}$ algebra.

In the previous section we introduced the action
\begin{equation}
S = \kappa {\cal S}^-[w] + \frac{1}{\pi x} \int \mbox{str}
 \left\{ \mu W \right\} .
\label{Wactie}
\end{equation}
It describes a fully constrained WZW model, of which the highest
weight
currents $W$ are coupled to chiral $ \cal W$ gravitational
(lowest $sl(2)$ weight) field $\mu$. The
currents transform under $\cal W$ transformations as in
eq.(\ref{delWhw}).

Consider the variation of the action $S$:
\begin{equation}
\delta_{\eta} S = \frac{1}{\pi x} \int \text{str}
  \left\{ -\bar\partial \eta\cdot W + \delta_{\eta} \mu \cdot W +
   \frac{\kappa}{2} \mu D\left[ 2W/\kappa\right]
   \frac{1}{I\left[ 2W/\kappa\right]} \eta \right\} .
\label{variat}
\end{equation}
It can be derived by using
\begin{equation}
\delta_X k {\cal S}^-[g] = \frac{-1}{\pi x} \int \, \mbox{str}
\left\{ \bar\partial X\, .\, J \right\} \,\,
\label{varSmin}
\end{equation}
and the fact that $\eta$ is of lowest
weight. The $W$-independent part of the variation (\ref{variat})
reads
\begin{equation}
\left. \left(\delta_{\eta} S \right) \right|_{W=0} = \frac{1}{\pi x}
\int
 \text{str} \left\{ \frac{\kappa}{2} \mu
 \frac{\partial}{1 -L \partial} \eta \right\} \,,\label{classanom}
\end{equation}
which can not be canceled by the $\delta_{\eta} \mu$ term.
This shows that, already at the level of the classical realization, we
have to face the central extension terms, which in some treatments
appear only at the quantum level. Although this forces one to arrange
for a cancellation also at this classical level, it is in fact a
blessing in disguise, since exactly the same cancellation mechanism
turns out to suffice for the quantum treatment.

\section{ Non-critical $\cal W$ string models}\label{non-critW}

In this section, we will lift the obstruction to the $\cal W$
invariance of the classical realization by introducing, besides the
matter sector, also the Liouville sector. Then, adding ghosts, we show
how this can be used to deduce the BRST charge of \cite{W3BRST} for
the combined system.

\subsection{The $\cal W$ invariant action}\label{non-critW:action}

The $\cal W$ transformations {\em can} be gauged if we introduce
{\em two}
WZW models, which we call ``matter"(M) and ``gravity"(G)
respectively. For
convenience we introduce the following notations
\begin{eqnarray}
D_M &\equiv& D\left[ 2W_M/\kappa_M\right] \nonumber\\
I_M &\equiv& I\left[ 2W_M/\kappa_M \right] = 1 - LD_M \,.
\label{notatie1}
\end{eqnarray}
Later on we will also need
the conjugated operator $I^+_M$, which is
defined by
\begin{equation}
I^+_M \equiv 1 - D_M L .  \label{notatie2}
\end{equation}
All these definitions of course apply, mutatis mutandis, for the
gravitational sector as well.

Our action at this stage is
\begin{equation}
S_{M+G} = \kappa_M {\cal S}^-[w_M] + \kappa_G {\cal S}^-[w_G]
+\frac{1}{\pi x} \int
\mbox{str} \left\{ \mu ( W_M + W_G) \right\}.
\label{mettwee}
\end{equation}
{}From eq.(\ref{classanom}) it is seen that the obstruction to
invariance
is lifted if the levels of the matter and gravity sector add up to
zero
\begin{equation}
\kappa_M + \kappa_G =0 .
\end{equation}
Using this relation, it remains to be checked that the last term in
the resulting variation of the action eq.(\ref{mettwee}),
\begin{eqnarray}
\delta_{\eta} S_{M+G} &=& \frac{1}{\pi x} \int \text{str}
 \left\{ -\bar\partial \eta\cdot (W_M + W_G) +
 \delta_{\eta} \mu \cdot (W_M + W_G) \right. \nonumber \\
&& \phantom{\frac{1}{\pi x} \int \text{str} }
 \left.+ \frac{\kappa_M}{2} \mu
 \left( D_M \frac{1}{I_M} - D_G \frac{1}{I_G} \right) \eta \right\}
\, ,
\end{eqnarray}
is proportional to $W_M+W_G$. Indeed, we find that
\begin{equation}
{ \int \text{str} \left\{ \frac{\kappa_M}{2} \mu
\left( \frac{1}{I^+_M} D_M - D_G \frac{1}{I_G} \right) \eta \right\} }
= - \int \text{str} \left\{ \mu \frac{1}{I^+_M} \mbox{ad} (W_M + W_G)
\frac{1}{I_G} \eta \right\} \, ,
\end{equation}
so that $S_{M+G}$ is invariant under ${\cal W}$ transformations if we
define
\begin{equation}
\delta_{\eta} \mu = \bar{\partial} \eta -
 \Pi_{\text{lw}} \left. \mbox{ad}
 \left( \frac{1}{I_M} \mu \right)
 \frac{1}{I_G} \eta \right|_{M,G} .
\label{deltamu}
\end{equation}
There is some arbitrariness in this choice.
The symbol $\left. \right|_{M,G}$
indicates that we have chosen an additive
$M$-$G$ symmetrization of the transformation law for $\mu$.
Explicitly,
$\left. F(M,G)\right|_{M,G}=\frac{1}{2}\left\{F(M,G)+F(G,M)\right\}$,
and  $D_{M,G} = \frac{1}{2} (D_M + D_G)$.

The gauge fixing of the action (\ref{mettwee}) is a  non-trivial
problem. It can for instance be checked that the ${\cal W}$ gauge
algebra
in general only closes modulo $W_M + W_G$ terms.
This will cause higher ghost interaction terms in the gauge fixed
theory.
In section \ref{BV} we will treat the derivation of these terms in
some
detail using the formalism of Batalin and Vilkovisky, which is
eminently
suited to master these complications. At the moment we only present
the lowest order terms explicitly:
\begin{eqnarray}
S_{\text{gf}} &=& \kappa_M {\cal S}^-[w_M] + \kappa_G {\cal S}^-[w_G]
 + \frac{1}{\pi
x}\int \mbox{str} \left\{ b \bar{\partial} c \right\}
\label{hiermee}
\\
&& + \frac{1}{\pi x}\int \mbox{str}\left\{ \widehat{\mu} \, \left( W_M+W_G +
\left. \frac{1}{I^+_M} ad(b) \frac{1}{I_G} c \right|_{M,G}
\right) + \mbox{more ghosts} \right\} . \nonumber
\end{eqnarray}

\subsection{The BRST charge of non-critical ${\cal W}_3$ strings}
\label{non-critW:BRST}
The BRST charge for ${\cal W}_3$ gravity can be read off from the
gauge
fixed action (\ref{hiermee}). Let us explain why this is the case.
The
background field $\widehat{\mu}$ that was introduced during the gauge
fixing of
the ${\cal W}$ symmetry of our model, is in fact nothing but the
antifield $b^*$
for the antighost $b$. But this means that operator that couples to
the
field $\widehat{\mu}$ is nothing but the BRST variation of $b$. The
BRST
transformation of $b$ splits into three distincts pieces
\begin{equation}
\delta_{\text{BRS}} b \sim W_M + W_G + W_{\text{gh}} ,
\label{sumcurr}
\end{equation}
where the ghost current $W_{\text{gh}}$ is given by
\begin{equation}
W_{\text{gh}} =
 \left. \Pi_{\text{hw}}\frac{1}{I^+_M} ad(b) \frac{1}{I_G} c
 \right|_{M,G}
+ \cdots.
\label{decurr}
\end{equation}
On the other hand we know that the BRST charge $Q$ when acting on
$b$,
generates the BRST transformation (\ref{sumcurr}), so $Q$
can easily be constructed once the $W$ currents are known.

For the case of ${\cal W}_3$ gravity we evaluate the ghost current
$W_{\text{gh}}$
explicitly. It contains terms quadratic in the ghosts only. If we
parametrize
\begin{equation}
W_\alpha = \left( \begin{array}{ccc}
               0& \frac{1}{4} T_\alpha &\frac{1}{2} W_{3,\alpha} \\
               0&0& \frac{1}{4} T_\alpha  \\
               0&0&0 \end{array} \right)
\qquad \qquad \mbox{for} \,\,\alpha  = M, G, \mbox{gh}
\label{w3-not1}
\end{equation}
and
\begin{equation}
b = \left( \begin{array}{ccc}
               0&\frac{1}{4} b_1 & \frac{1}{2} b_2\\
               0&0&\frac{1}{4} b_1\\
               0&0&0 \end{array} \right)
\qquad
c = \left( \begin{array}{ccc}
               0&0&0\\
               c_1&0&0\\
               c_2&c_1&0 \end{array} \right)
\label{w3-notations}
\end{equation}
we find that
\begin{eqnarray}
T_{\text{gh}} &=& -2 b_1\partial c_1 -\partial b_1\cdot c_1 -
3b_2\partial c_2-2\partial
b_2 \cdot c_2 \nonumber \\[2mm]
W_{3,\text{gh}} &=& -3b_2\partial c_1 - \partial b_2\cdot c_1
-\frac{2}{3\kappa_M} b_1\partial c_2 \cdot\left( T_M - T_G \right) \\
&& -\frac{1}{3\kappa_M} \partial b_1\cdot c_2 \left( T_M - T_G
\right)
 - \frac{1}{3\kappa_M} b_1 c_2 \left( \partial T_M - \partial T_G
\right) \nonumber \\[2mm]
&& +\frac{1}{12} \left\{ 10 b_1\partial ^3 c_2 + 15 \partial
b_1\cdot\partial^2c_2
+ 9\partial^2b_1\cdot\partial c_2 + 2 \partial^3b_1 \cdot c_2
\right\}
\label{Teee}
\end{eqnarray}
To compare our result with the currents $T_{\text{gh}}$ and
$W_{3,\text{gh}}$
that were obtained in \cite{W3BRST} we introduce rescaled
spin 3 ghosts $b'_2$ and $c'_2$ :
\begin{equation}
b_2 = \frac{1}{\sqrt{\kappa_M}} b'_2  \qquad\qquad c_2=
\sqrt{\kappa_M}
c'_2 \end{equation}
To make this rescaling into a canonical operation we also redefine
the
antifields of the ghosts. It is then very natural to rescale the
background field $\widehat{\mu}_3$, and the $W_3$ currents as well
\begin{equation}
\widehat{\mu}_3 = \sqrt{\kappa_M} \widehat{\mu}'_3  \qquad\qquad
W_{3,\alpha} = \frac{1}{\sqrt{\kappa_M}} W_{3,q\alpha}'   .
\end{equation}
The rescaled ghost current $W_{3,\text{gh}}'$ reads (dropping the
primes)
\begin{eqnarray}
W_{3,\text{gh}} &=& -3b_2\partial c_1 - \partial b_2\cdot c_1
-\frac{2}{3} b_1\partial c_2 \cdot\left( T_M - T_G \right)
\label{Weee}
\\
&& -\frac{1}{3} \partial b_1\cdot c_2 \left( T_M - T_G \right)
 - \frac{1}{3} b_1 c_2 \left( \partial T_M - \partial T_G \right)
\nonumber \\[2mm]
&& +\frac{\kappa_M}{12} \left\{ 10 b_1\partial ^3 c_2 + 15 \partial
b_1\cdot\partial^2c_2
+ 9\partial^2b_1\cdot\partial c_2 + 2 \partial^3b_1 \cdot c_2
\right\} \nonumber .
\end{eqnarray}
Now we comment on the transition to quantum theory. There is a
general
formula \cite{Zfac} for arbitrary DS reductions,
\begin{equation}
c = \frac{1}{2} c_{\text{crit}} -
 \frac{(d_B-d_F)\tilde h }{\kappa + \tilde h }
 - 6y (\kappa +\tilde h ) .
\label{cpretty}
\end{equation}
where $c_{\text{crit}}$ is the critical value of the central charge
for
the ${\cal W}$ algebra under consideration,
$d_B$ and $d_F$ count the number of bosonic and fermionic
generators in the Lie algebra $\bar g$, and
$y$ is the index of embedding of $sl(2)$ in $\bar g$.
The values of these characteristic numbers can be computed with
simple
counting formulas \cite{Zfac}.
For the case at hand, the DS reduction of the ${\cal W}_3$ algebra
proceeds
via the principal embedding  of
$sl(2)$ in $sl(3)$ (so $d_B=8,d_F=0$), and the $sl(3)$ algebra
branches
into an $sl(2)$ spin $j=1$ and $j=2$ representation. The values
$c_{\text{crit}}=100$ and $y=4$ follow.
In the limit of large central charges (which in our case corresponds
to the classical limit)
$-24 \kappa_M = c_M$, as is clear from (\ref{cpretty}).
We may write the factor
\begin{equation}
\frac{\kappa_M}{12} = -\frac{1}{90} \frac{5c_M}{16} =
-\frac{1}{90\beta^0_M} \,. \end{equation}
Upon quantization this factor, and only this factor, must be
renormalized
\begin{equation}
-\frac{1}{90\beta^0_M} \mapsto \frac{17 \beta_M -1}{90\beta_M}
\,\,\,\mbox{with}\,\,\, \beta_M = \frac{16}{22+5c_M} \,,
\end{equation}
leading immediately to the nilpotent BRST charge \cite{BLNW,W3BRST}
\begin{equation}
Q_{\text{non-crit},W_3} = \oint \frac{dz}{2 \pi i}
c_1 \left( T_M + T_G + \frac{1}{2} T_{\text{gh}} \right)
+ c_2 \left( W_{3,M} + W_{3,G} + \frac{1}{2} W_{3,\text{gh}} \right)
\end{equation}
This may be compared with the procedure in \cite{W3BRST}, where the
same
final result was obtained only after adding additional terms to a
classical charge.
The reader wil have noticed that in the present treatment
the BRST charge follows almost
automatically from the equation (\ref{decurr}). Once the classical
ghost currents of eqs. (\ref{Teee}) and (\ref{Weee}) have been
derived,
one can obtain the quantum currents by a simple renormalization of
one
factor in front of the classical terms. In this respect the
realization of
the ${\cal W}_3$ algebra via WZW models, proves to be superior to the
realization in terms of scalar fields which was used in
\cite{W3BRST}.
In the classical analysis of \cite{W3BRST} the term proportional to
$\kappa_M\sim c_M$ in (\ref{Weee}) was absent,
and arose at the
quantum level from counterterms. Clearly, using WZW models one
already has a
non-zero
central charge at the classical level, so that the transition to the
quantum theory can proceed in a very gentle way.

\section{Gauge Fixing}\label{BV}
In this section we treat more thoroughly the questions related to
gauge
fixing, both for the Drinfeld-Sokolov symmetry and for the $\cal W$
symmetries. For the DS symmetry we present a realization of the
gauging
that, at the expense of introducing extra Lagrange multipliers,
succeeds
in closing the algebra of the transformations.
As a result the gauge fixing procedure simplifies, and although one
could
dispense with the full Batalin-Vilkovisky treatment\footnote{A review
of the Batalin-Vilkovisky formalism can be found in \cite{boek}.}, we
nevertheless
phrase it in that language for uniformity.
For the $\cal W$ symmetries our treatment does not (at least not
automatically) lead to such a simplified algebra. Because the
symmetries
close only modulo field equations, it is expedient to use the BV
treatment to take this into account. We will not succeed in deducing
all
order (in antifields) expressions for arbitrary DS reductions, but at
the
end we will illustrate the general procedure by deriving the relevant
expression for the ${\cal W}_3$ case.

\subsection{The Drinfeld-Sokolov symmetry}\label{BV:DS}
The relevant information concerning the Drinfeld-Sokolov symmetries,
which
are quite conventional gauge symmetries,
are encoded by adding the antifield-dependent terms
\begin{equation}
S_* = \frac{1}{\pi x} \int \mbox{str} \left\{- \frac{\kappa}{2}
\tilde J ^* D [
2\tilde J  /\kappa ] c_{\text{DS} } + \frac{1}{2} c^*_{\text{DS}}
\mbox{ad}
\left( c_{\text{DS} }\right) c_{\text{DS} }\right\} ,
\end{equation}
where $\tilde J$ is given in eq.(\ref{vormtJ}) and
$c_{\text{DS}} \in \Pi_{> 0} \text{\euler g}$.
The extended action $S_{1,\text{ext}}=S+S_*$, with $S$ from
eq.(\ref{actiesplit}), is a cornerstone of the
Batalin-Vilkovisky treatment. Gauge invariance is expressed through
the classical master equation
$(S_{1,\text{ext}},S_{1,\text{ext}})=0$.
The term in the extended action proportional to
$c^*_{\text{DS}}$ expresses the closure of the DS gauge algebra. The
particular form of this $c^*_{\text{DS}}$ dependent term is typical
for
non-abelian gauge theories.

To proceed, we now add a (cohomologically) trivial system,
with the extended action
\begin{eqnarray}
S_{\text{triv}} &=& \kappa{\cal S}^-[g ] - \kappa{\cal S}^-[\tilde g
] + \frac{1}{\pi x}
\int \mbox{str} \left\{ A (J-\tilde J  ) \right\} \nonumber \\[2mm]
&&- \frac{1}{\pi x} \int \mbox{str} \left\{ \frac{\kappa}{2} J^* D
\left[
2J/\kappa \right] c_{\text{DS} } + A^* \bar D \left[ A \right]
c_{\text{DS} }
\right\}\, .
\label{trivwzwalg}
\end{eqnarray}
The extra variables introduced here are a Lie algebra valued Lagrange
multiplier $A$, and an extra
current $J$ which is completely unconstrained. The action
is   trivial in the antibracket sense.
The addition of this extra trivial system allows us to ``unconstrain"
the currents on which the DS transformations are acting, achieving in
this
way a decoupling of the constraints and the gauge transformations.
This is
the basic reason why we succeed in obtaining a {\em closed} algebra,
which, upon elimination of the trivial systems (by integrating out
the
Lagrange multipliers and putting their antifields to zero), goes over
into
the open algebra computed in \cite{W3allord,Zfac}. This we now show.
We split the full
Lagrange multiplier $A$ and its antifield into two parts:
\begin{eqnarray}
A = A_{\text{DS} } + A_{\text{ident}}
      &\quad\mbox{with}\quad&
    A_{\text{DS} } \in\Pi_{> 0} \text{\euler g}     \, ; \quad
    A_{\text{ident}} \in\Pi_{\leq 0} \text{\euler g}
 \nonumber \\[2mm]
A^* = A^*_{\text{DS} } + A^*_{\text{ident}}
        &\quad\mbox{with}\quad&
      A^*_{\text{DS} } \in\Pi_{< 0} \text{\euler g} \, ; \quad
      A^*_{\text{ident}} \in\Pi_{\geq 0} \text{\euler g} .
\end{eqnarray}
The Lagrange multipliers in $A_{\text{DS}}$ are precisely the ones
that
impose the Drinfeld-Sokolov constraints, bringing the current $J$
into
the $\tilde J $ form. We keep these Lagrange multipliers manifest in
the
action.
The multipliers in $A_{\text{ident}}$ identify the free components
$J^{\geq 0}$ which are contained in $\tilde J $, with the
corresponding
components in $J$.
We will implement this identification, by integrating explicitly over
$A_{\text{ident}}$ and over $J^{\geq 0}$.
To this end we rewrite the extended action $S + S_* +
S_{\text{triv}}$ as
\begin{eqnarray}
S_{2,\text{ext}} &=&
 \kappa {\cal S}^-[g] + \frac{\kappa}{4\pi x} \int \mbox{str}
 \left\{ [\tau,e_-] \bar{\partial} \tau \right\} +
 \frac{1}{\pi x} \int \mbox{str} \left\{ \mu W(\tilde J ) \right\}
 \nonumber \\[2mm]
&&+ \frac{1}{\pi x} \int \mbox{str}
 \left\{
  \left( A_{\text{DS} } + A_{\text{ident}}\right)
  \left( J - \mbox{ad}
   \left( A^*_{\text{DS} } +A^*_{\text{ident}}\right)
   c_{\text{DS} } -\tilde J  \right) \right\}
 \nonumber \\[2mm]
&& + \frac{1}{\pi x} \int \mbox{str}
 \left\{ - A^*_{\text{DS} } \bar{\partial} c_{\text{DS} }-
  \frac{\kappa}{2} \tilde J ^* D [ 2\tilde J /\kappa]
   c_{\text{DS} } -\frac{\kappa}{2} J^* D
  \left[ 2J/\kappa\right] c_{\text{DS} } \right\}
 \nonumber \\[2mm]
&&+ \frac{1}{\pi x} \int \mbox{str}
 \left\{ \frac{1}{2} c^*_{\text{DS}} \mbox{ad}
  \left( c_{\text{DS} }\right)
  c_{\text{DS} }\right\}\, .
\end{eqnarray}
Next we introduce the shifted current
\begin{equation}
\check{J} = J - \mbox{ad}\left( A^*_{\text{DS} } \right) c_{\text{DS}
}
\end{equation}
and now eliminate the $A_{\text{ident}}$ and $J^{\geq 0}$ fields,
with their
corresponding antifields. This leads to the extended action
\begin{eqnarray}
S_{\text{ext}} &=&
 \kappa {\cal S}^-[g] + \frac{\kappa}{4\pi x} \int \mbox{str}
 \left\{ [\tau,e_-] \bar{\partial} \tau \right\} +
 \frac{1}{\pi x} \int \mbox{str}
 \left\{ \mu W(\check{J}^{\geq 0},\tau) \right\}
 \nonumber \\[2mm]
&&+ \frac{1}{\pi x} \int \mbox{str}
 \left\{ A_{\text{DS} }
  \left( \check{J}^{<0} - \frac{\kappa}{2} e_- -
   \frac{\kappa}{2} [\tau,e_- ]\right)
  - A^*_{\text{DS} } \bar{\partial} c_{\text{DS} } \right\}
 \nonumber \\[2mm]
&&+ \frac{1}{\pi x} \int \mbox{str}
 \left\{ -\frac{\kappa}{2} J^* D \left[ 2J/\kappa\right] c_{\text{DS}
}
  - \tau^* \Pi_{+1/2} c_{\text{DS} } \right\}
 \nonumber \\[2mm]
&&+ \frac{1}{\pi x} \int \mbox{str}
 \left\{ \frac{1}{2} c^*_{\text{DS}}
  \mbox{ad} \left( c_{\text{DS} }\right) c_{\text{DS} }\right\} .
\label{finalextact}
\end{eqnarray}
Notice that this action may contain terms with multiple antifields
$A^*_{\text{DS} }$, due
to the appearance of shifted currents in the gauge invariant
polynomials
$W(\check{J}^{\geq 0},\tau )$. If this happens, this is a
manifestation of the
non-closure of the gauge algebra, that belongs to the DS invariant
classical action
$S_{\text{cl}}=S_{\text{ext}}[A^*=J^*=c^*=\tau^*=0]$.
It is precisely this classical action $S_{\text{cl}}$ that was used
in \cite{W3allord} in the case of ${\cal W}_3$
gravity, and in \cite{Zfac} in the case of $SO(N)$ supergravities, as
a starting point for a direct construction of the BV-extended
action. The existence of non-closure terms made this construction
rather cumbersome, but as we showed here, this can be avoided by
introducing a redundant set of Lagrange multipliers
$A = A_{\text{DS} } + A_{\text{ident}}$,
which keeps the gauge algebra closed. The BV extended action can be
constructed easily in this extended space of variables,
and be reduced afterwards.

The gauge fixing of the DS symmetry in the extended action
(\ref{finalextact}) can now be simply achieved by putting
$A_{\text{DS} } = \widehat{A}_{\text{DS} }= b^*_{\text{DS} }$
and $A^*_{\text{DS} } = - b_{\text{DS} }$, a transformation of
variables
canonical in the antibracket.
This is one of the gauges used in \cite{Zfac}.
If we keep the dependence on  $\widehat{A}_{\text{DS} }$,
so that the reader may still transit to the other gauge used in
\cite{Zfac} if (s)he wants, we find
\begin{eqnarray}
S_{\text{gf}} &=& \kappa {\cal S}^-[g] + \frac{\kappa}{4\pi x} \int
\mbox{str}
\left\{
[\tau,e_-] \bar{\partial} \tau \right\} + \frac{1}{\pi x} \int
\mbox{str}\left\{
b_{\text{DS} } \bar{\partial} c_{\text{DS} } \right\} \nonumber
\\[2mm]
&&+ \frac{1}{\pi x} \int \mbox{str}\left\{ \mu W(\check{J}^{\geq
0},\tau) +
\widehat{A}_{\text{DS} } \left( \check{J}^{<0} -
\frac{\kappa}{2} e_- - \frac{\kappa}{2} [\tau ,e_- ]\right) \right\}
\nonumber \\[2mm]
\label{finalgfact}
\end{eqnarray}
where, apart from the kinetic term, the ghost dependence is
through the shifted current
\begin{equation}
\check{J} = J + \mbox{ad}\left( b_{\text{DS} } \right) c_{\text{DS} }
{}.
\end{equation}
It should be remarked that the BRST transformation rules of the
fields in the gauge fixed action do not depend on the sources
$\mu$. From this we learn that the DS invariant polynomials
computed from eq.(\ref{finalpol}), have become {\em BRST
invariant} polynomials through the replacement of the currents
$J^{\geq
0}$ by the shifted $\check{J}^{\geq 0}$.

\subsection{The ${\cal W}$ gauge symmetry}\label{BV:W}

We propose to start from an unconstrained system of coupled WZW
models, for which the extended action can be obtained more easily.
Using only canonical methods (with respect to the antibracket) we
then implement the various constraints, necessary to bring the WZW
models in the highest weight form.

The starting point is
\begin{eqnarray}
S_0 &=&
 \kappa_M {\cal S}^-[g_M] + \kappa_G {\cal S}^-[g_G] +
 \frac{1}{\pi x}\int str\left\{ A \, \left( J_M + J_G \right)
\right\}
 \nonumber \\[2mm]
&& - \frac{1}{\pi x}\int str
 \left\{ A^* \, \bar D[A] C -
  \frac{1}{2} C^* \mbox{ad}\left( C \right) C \right\}
 \nonumber \\[2mm]
&& - \frac{1}{\pi x}\int str
 \left\{ \frac{\kappa_M}{2} J^*_M D[{2J_M/\kappa}] C +
  \frac{\kappa_G}{2} J^*_G D[{2J_G/\kappa}] C \right\} ,
\end{eqnarray}
where all the fields take values in the entire Lie algebra, and the
covariant derivatives involve at the moment
unconstrained currents $J_M$ and $J_G$.
One may notice that we are treating the currents $J$ as basic
variables, rather than the group elements $g$: this simplifies
the calculations, but should not influence the results.
One can read off the gauge (or BRST) transformations from the terms
with starred fields. The gauge invariance (i.e. the BV master
equation) can be checked explicitly if  $\kappa_M+\kappa_G=0$.
It can also be seen by
parametrizing $A=h^{-1}\bar\partial h$, and rewriting the first line,
with the help of the Polyakov-Wiegmann formula eq.(\ref{PolWform}),
as the sum of two (separately invariant) WZW actions
$\kappa_M {\cal S}^-[h g_M] + \kappa_G {\cal S}^-[h g_G]$:
the condition $\kappa_M+\kappa_G=0$ eliminates the additional
${\cal S}^-[h]$ terms.
The gauge field $A$ acts as a Lagrange multiplier imposing
$J_M + J_G - \mbox{ad}(A^*)C = 0$.
The antifield dependence of this constraint can be
absorbed into a redefinition of the currents $J_M$, $J_G$. We
implement this redefinition by performing the canonical
transformation generated by
\begin{equation}
F= {{\bf 1}} - str
 \left\{ \frac{1}{2} \left( J_M^{'*} + J_G^{'*} \right)
  \mbox{ad}\left( A^{'*} \right) C \right\} .
\end{equation}
Dropping the primes, it leads to the following extended action:
\begin{eqnarray}
S_1 &=& \kappa_M {\cal S}^-[h_M] + \kappa_G {\cal S}^-[h_G]
 + \frac{1}{\pi x}\int str\left\{ A \, \left( J_M + J_G \right)
\right\}
\nonumber \\[2mm]
&& - \frac{1}{\pi x}\int str\left\{ A^* \, \bar{\partial} C
-\frac{1}{2} C^*
\mbox{ad}\left( C \right) C\right\} \nonumber \\[2mm]
&& - \frac{1}{\pi x}\int str\left\{ \frac{\kappa_M}{2} J^*_M D_{J_{av}}
C + \frac{\kappa_G}{2} J^*_G D_{J_{av}} C
\right\} ,
\end{eqnarray}
where the covariant derivative
$D_{J_{av}}=D[J_M/\kappa_M+J_G/\kappa_G]$ involves a current
that averages over matter and gravitational sectors, and
the group elements $h_\alpha$, for $\alpha \in \{ M, G\}$,
are defined through
\begin{equation}
\frac{\kappa_\alpha }{2} \partial h_\alpha h^{-1}_\alpha = J_\alpha +
\frac{1}{2}
\mbox{ad}(A^*) C .
\end{equation}
The next step is to split the gauge field $A$ into pieces, say $A =
\overline{\Pi}_{\text{lw}} A + \mu \equiv \overline{A} + \mu$, and
accordingly
$A^* = \overline{\Pi}_{\text{hw}} A^* + \mu^* \equiv \overline{A}^* +
\mu^*$.\footnote{By definition $\overline{\Pi}_{\text{lw}} = 1 -
\Pi_{\text{lw}}$.} It is clear that the $\overline{A}$
field imposes the condition
$ \overline{\Pi}_{\text{hw}} \left( J_M +J_G \right) = 0$.
To achieve our aim of constraining both currents in the
Drinfeld-Sokolov way, we need an extra condition. The gauge
freedom allows us to impose such a condition. We choose to impose
it in a $M\leftrightarrow G$ symmetric way:
the condition
$\overline{\Pi}_{\text{hw}}
 \left( \frac{J_M}{\kappa_M} + \frac{J_G}{\kappa_G} - e_-\right) = 0$
precisely brings the currents $J_M$ and $J_G$ in the desired highest
weight form. In the Batalin-Vilkovisky scheme we may implement that
constraint by first adding the following trivial system to
the action:
\begin{equation}
S_{\text{triv}} = \frac{1}{\pi x}\int str\left\{ \rho^* \, \lambda
\right\}
, \end{equation}
where $\lambda,\rho \in \overline{\Pi}_{\text{lw}} \text{\euler g}$. Then
we perform
the canonical transformation with generator
\begin{equation}
F= {{\bf 1}} + str \left\{ \rho \, \overline{\Pi}_{\text{hw}} \left(
\frac{J_M}{\kappa_M} + \frac{J_G}{\kappa_G} - e_-  \right) \right\}.
\end{equation}
The resulting extended action reads
\begin{eqnarray}
S_2 &=& \kappa_M {\cal S}^-[h_M] + \kappa_G {\cal S}^-[h_G] +
 \frac{1}{\pi x} \int
 str\left\{ \mu \, \left( V_M + V_G \right) \right\}
\nonumber \\[2mm]
&& -\frac{1}{\pi x}\int
 str\left\{ \left( \overline{A}^* + \mu^* \right)\, \bar{\partial} C
 -\frac{1}{2} C^* ad\left( C \right) C
 - \overline{A} \, \overline{\Pi}_{\text{hw}}
 \left( J_M + J_G \right) \right\}
\nonumber \\[2mm]
&& - \frac{1}{\pi x}\int str\left\{
 \frac{\kappa_M}{2} J^*_M D_{J_{av}} C
 + \frac{\kappa_G}{2} J^*_G D_{J_{av}} C \right\}
\nonumber \\[2mm]
&& - \frac{1}{\pi x}\int str\left\{ \rho D_{J_{av}} C
 - \lambda \, \overline{\Pi}_{\text{hw}}
 \left( \frac{J_M}{\kappa_M} + \frac{J_G}{\kappa_G}
       - e_- + \rho^* \right) \right\} ,
\end{eqnarray}
where the currents $V_\alpha$ are the highest weight
components of the $J_\alpha$'s. Now we eliminate the variables
$\left\{
\overline{A}, \lambda,
\overline{\Pi}_{\text{hw}} J_M, \overline{\Pi}_{\text{hw}} J_G
\right\}$, and
find that
\begin{eqnarray}
S_3 &=& \kappa_M {\cal S}^-[f_M] + \kappa_G {\cal S}^-[f_G] +
\frac{1}{\pi x}\int
str\left\{ \mu \, \left( V_M + V_G \right) \right\} \nonumber \\[2mm]
&& - \frac{1}{\pi x}\int str\left\{ \mu^* \, \bar{\partial}
C-\frac{1}{2} C^* \mbox{ad}\left( C \right) C + \rho \left( D_{V_{av}} -
E_- +\mbox{ad}(\rho^*) \right) C\right\} \nonumber \\[2mm]
&& - \frac{1}{\pi x}\int str\left\{ \left. \kappa_M V^*_M
\left( D_{V_{av}} - E_- +\mbox{ad}(\rho^*) \right) C \right|_{M,G}
\right\}  \,.
\end{eqnarray}
The group elements $f_\alpha$ are given by
\begin{equation}
\frac{\kappa_\alpha }{2} \partial f_\alpha f^{-1}_\alpha =
\frac{\kappa_\alpha}{2} (e_- - \rho^*) + V_\alpha + \frac{1}{2}
\mbox{ad}(\mu^*)
C . \end{equation}

The fields $\left\{ \rho, \overline{\Pi}_{\text{lw}} C\right\}$ also
form a ``trivial" pair of variables, albeit in a more subtle way.
Indeed, the equation of motion of $\rho$ evaluated in the point
$\rho^* = 0$ is equivalent to eq.(\ref{varww}). The structure of
this equation is such that all the $\overline{\Pi}_{\text{lw}}C$
fields can be exactly solved for, yielding
\begin{equation}
C \rightarrow \frac{1}{ I_{V_{av}}} c ,
\end{equation}
where $c$ denotes the lowest weight part of the original ghost field
$C$ and $I_{V_{av}}$ is defined in terms of the average current as
$I_{V_{av}}=1-LD_{V_{av}}$. In doing so we find the action
\begin{eqnarray}
S_4 &=& \kappa_M {\cal S}^-[v_M] + \kappa_G {\cal S}^-[v_G] +
 \frac{1}{\pi x}\int str
 \left\{ \mu \,
  \left( V_M + V_G \right)
 \right\} \nonumber \\[2mm]
&&- \frac{1}{\pi x}\int str
     \left\{ \mu^* \, \bar{\partial} c
      -\frac{1}{2} c^* \text{ad}
      \left( \frac{1}{ I_{V_{av}}} c \right)
      \frac{1}{ I_{V_{av}}} c
     \right\}
\label{oef}
\\
&&- \frac{1}{\pi x}\int str\left\{ \frac{\kappa_M}{2} V^*_M
D_{V_{av}}  \frac{1}{ I_{V_{av}}} c+
\frac{\kappa_G}{2} V^*_G D_{V_{av}} \frac{1}{ I_{V_{av}}} c \right\} ,
\nonumber
\end{eqnarray}
with
\begin{equation}
\frac{\kappa_\alpha }{2} \partial v_\alpha v^{-1}_\alpha =
\frac{\kappa_\alpha}{2} e_-  + V_\alpha + \frac{1}{2}
\mbox{ad}(\mu^*) \frac{1}{ I_{V_{av}}} c .   \label{shiftedJ}
\end{equation}
The last term in eq.(\ref{shiftedJ}) will be called the ghost
current $J_{gh}$.
We may replace $\mu^*$ by (minus) the antighost $b$, and put
$\mu$ equal to a background value.
The action (\ref{oef}) then becomes the gauge fixed action.

This expression, albeit not very transparant, is valid to all orders
in the ghost fields. The ghost field dependence is partly explicit,
but also implicit in the WZW functionals, where it enters through the
definition of the group elements $v_\alpha$ in eq.(\ref{shiftedJ}).
We now investigate how to make this dependence more explicit.
Although at present we can not give the end result in general,
the following constitutes a constructive procedure.
We will explicitize the ghost dependence in a
specific case, namely the reduction of $sl(3)$ to the
${\cal W}_3$ algebra,
which also served as an example in section~\ref{non-critW:BRST}.

The first step is, to disentangle the dependence in the WZW actions.
To this end, in the same spirit as in section~\ref{Wmat:DS},
we factorise $v_\alpha = e^{-\gamma_\alpha} w_\alpha$.
where $w_\alpha$ is such that  the current
$\frac{\kappa_\alpha }{2} \partial w_\alpha w^{-1}_\alpha$
is in the highest weight form
$\frac{\kappa_\alpha}{2} e_- + W_\alpha$.
To obtain this form, we follow the same method as in
section~\ref{Wmat:DS}. Note however that the right hand side
of eq.(\ref{shiftedJ}) is not restricted to
non-negative $e_0$-grading, due to the ghost contribution.
It is not obvious from the group property that such a heighest
weight gauge can be reached.
Proceeding nevertheless in the same manner,
we put $\gamma_\alpha =\sum_{n\ge 1}\gamma_\alpha^{(n)}$ and
$W_\alpha =\sum_{n \ge 0}W_\alpha^{(n)}$, where the expansion now is not
in the full current (as in section~\ref{Wmat:DS} ), but in the
deviation from the highest weight form, namely  the ghost current
in eq.(\ref{shiftedJ}). Consequently, the successive terms in this
expansion will be sums of products of two, four, six, etc. ghost
fields.
In addition we impose
$\gamma_\alpha \in \bar\Pi_{\text{lw}}\mbox{\euler g}$,
which guarantees that $LE_-\gamma_\alpha = \gamma_\alpha$.
Now an algorithm can be given to
construct  $\gamma_\alpha$ and $W_\alpha$ iteratively.
The recursive construction is:
\begin{eqnarray}
\gamma^{(0)}&=& 0 \quad;\quad\quad W^{(0)}_\alpha = V_\alpha
\nonumber \\
g^{(n)}&=&exp\left\{ \gamma_\alpha^{(0)} +\ldots
   +\gamma_\alpha^{(n-1)}\right\} \nonumber \\
X_\alpha^{(n)} &=&
 \frac{1}{I^+_\alpha} \,{\cal P}^{(n)}
  \left[g^{(n)}
   \left( e_- - D_\alpha + \frac{2}{\kappa_\alpha} J_{gh} \right)
     (g^{(n)})^{-1}
  \right] \,,
\nonumber  \\
\gamma_\alpha^{(n)} &=& L\, X_\alpha^{(n)} \,,\nonumber \\
W_\alpha^{(n)} &=& \frac{\kappa_\alpha}{2} \Pi_{hw}\,X_\alpha^{(n)}\,,
\qquad\qquad\qquad n\ge 1\,.
\end{eqnarray}
In these expressions, the derivatives are covariant derivatives, with
$V_\alpha$ (either $V_M$ or $V_L$) as gauge fields. These derivatives
are also used to construct $I_\alpha$ via eq.(\ref{defI}). Finally,
${\cal P}^{(n)}$ now denotes that only terms with products of $2n$
ghost fields are kept.
For concreteness, we list the first few terms
\begin{eqnarray}
\gamma_\alpha &=& \frac {2}{\kappa_\alpha} L \frac{1}{I^+_\alpha} J_{gh}
  - \frac{2}{\kappa_\alpha^2} L \frac{1}{I^+_\alpha}
  ad\left( J_{gh} + \Pi_{\text{hw}} \frac{1}{I^+_\alpha} J_{gh} \right)
  L\frac{1}{I^+_\alpha} J_{gh} + \cdots \nonumber \\
W_\alpha &=& V_\alpha + \Pi_{\text{hw}} \frac{1}{I^+_\alpha} J_{gh}
  - \frac{1}{\kappa_\alpha} \Pi_{\text{hw}} \frac{1}{I^+_\alpha}
  ad\left( J_{gh} + \Pi_{\text{hw}} \frac{1}{I^+_\alpha} J_{gh} \right)
  L\frac{1}{I^+_\alpha} J_{gh} + \cdots
\label{WinV}
\end{eqnarray}
Explicitly, for ${\cal W}_3$, we find the following relations in the
'matter' sector. We write down
the relations between the heighest weight fields $T$ and $W_3$
before ($V$) and after ($W$) the transformation with
$\gamma_\alpha$  with the (conventional) normalizations as in
eq.(\ref{w3-not1}):
\begin{eqnarray}
T_M(W)&=& T_M(V)+
 \{2\, b_1\,\partial c_1 + 3\, b_2\,\partial c_2
   +\partial b_1\, c_1 + 2\,\partial b_2\, c_2\}/ 2
\nonumber \\
&&\quad\quad\quad\quad\quad
+ 5\,{ b_1}\, {\partial b_1}\, { c_2}\, {\partial c_2}/ {48\,{ \kappa_M}}
\nonumber \\
W_{3,M}(W)&=& W_{3,M}(V)-
 \left\{
  -\partial b_2\, c_1/{2}  - 3\,b_2\,\partial c_1/{2}
 \right. \nonumber \\
&& \quad\quad\quad
  5\,b_1\, \partial^3 c_2/{12}
   + 5\,\partial b_1\,\partial^2 c_2/{8}
   + 3\,\partial^2b_1\, \partial c_2/{8}
   + \partial^3 b_1 \,c_2/{12}  \nonumber \\
&& \quad\quad\quad
     - T_{G}(V)\, b_1\, \partial c_2/{6\,\kappa_G}
   - T_{G}(V)\, \partial b_1\, c_2/{16\, \kappa_G}
     \nonumber \\
&& \quad\quad\quad
   -  { T_{M}(V)}\,b_1\,\partial c_2/{2\,{ \kappa_M}}
   - 13\,T_{M}(V)\,\partial b_1\, c_2/{48\,{ \kappa_M}}
\nonumber \\
&& \quad\quad\quad
    -{{5\,\partial{T_{G}(V)}\,{ b_1}\, { c_2}}/{48\,{ \kappa_G}}} -
   {{11\,{\partial T_{M}(V)}\, {b_1}\, { c_2}}/{48\,{ \kappa_M}}}\} \,
\nonumber \\
&& \quad\quad\quad
+\{
-24\,{ b_1}\, { b_2}\, { c_2}\, {\partial^2 c_2}
- 8\,{ b_1}\, {\partial b_2}\, { c_2}\, {\partial c_2}
- 24\,{\partial b_1}\, { b_2}\, { c_2}\, {\partial c_2}
\nonumber \\
&& \quad\quad\quad
+ 8\,{ b_1}\, {\partial b_1}\, { c_1}\, {\partial c_2}
- { b_1}\, {\partial b_1}\, {\partial c_1}\, { c_2}
+ 3\,{ b_1}\, {\partial^2 b_1}\, { c_1}\, { c_2}
\}/ {48\,{ \kappa_M}}
\label{W-V-ghosts}
\end{eqnarray}
For the 'Liouville' sector, the same relations obtain, mutatis
mutandis.
We do not write down, the corresponding
expansions for $\gamma_\alpha$.
We now construct the gauge fixed action
explicitly. The Polyakov-Wiegmann formula is used repeatedly
to extract the ghost dependence from the WZW functionals.
It turns out that all ghost contributions vanish,
for a variety of reasons: partial integration, highest weight
properties, Grassman algebra, and $\kappa_M+\kappa_G=0$.
The total currents $V_M+V_G$ in
eq.(\ref{oef}) can be obtained by inverting the relations
eqs.(\ref{W-V-ghosts}) above. Due to the presence of the four-ghost
terms, this is actually simpler for the sum than for $V_M$ and $V_G$
separately, by virtue of the relation $\kappa_M+\kappa_G=0$ which
causes the four-ghost terms to cancel. The result is that
$
V_M+V_G=W_M+W_G+W_{gh}
$,
with the ghost contributions given by eq.(\ref{Teee}). Thus we
fulfilled our promise in section~\ref{non-critW}. We emphasize
that the method used here was completely constructive.
Finally, the terms of eq.(\ref{oef}) involving antifields are
immaterial for the gauge fixed action (they  determine the final
constraint algebra), and need not be discussed here.

Having demonstrated the method, let us now comment on the general
situation. First of all, the gauge fixed action
that is implied by the eq.(\ref{oef}) has all the suitable variables
and symmetries. The dependence on the ghost fields, as emphasized,
is only given implicitly through
the shifted currents of eq.(\ref{shiftedJ}), making the
ghost Lagrangian paticularly untransparant. The strategy applied
above for ${\cal W}_3$ may be developed for the general case also,
but a
couple of possible obstructions to this straightforward line should
be mentioned. First, whereas in section~\ref{Wmat:DS} the finiteness
of the iteration in eq.(\ref{finalpol}) was guaranteed, for
eq.(\ref{W-V-ghosts}) we do not have such a proof,
although we do believe that there is no problem in this respect.
In particular, for reductions of Lie algebras (not superalgebras)
all ghosts are fermionic, and the finiteness of the expansions of
$\gamma_\alpha$ and $W_\alpha$ follows from dimensional arguments.
Perhaps more serious is the fact
that, in the general case, we have no reason to expect that the WZW
functionals with argument $e^{\gamma_\alpha}\, w_\alpha$
will always simplify
as for ${\cal W}_3$ above. In general, this could entail a non-standard
ghost Lagrangian, and quite possibly a further transformation
may be needed, of variables from the set
$
\{V_\alpha \,\, , \,\, \mbox{ghosts}\}
$
to
$\{W_\alpha
(V_\alpha, \text{ghosts})
\, , \,  \text{ghosts'} (V_\alpha, \text{ghosts}) \}
$
 that mixes the ghosts with the matter and gravity currents.
This transformation should be such that in the end the
redefined ghost fields  decouple from the WZW models.
Also, the inversion of the relations expressing the $W$ curents
in terms of the $V$ currents may be considerably more involved
in general.
We leave the treatment of these complications to the
future.

\section{Conclusions}
To recapitulate, we realized any $\cal W$ symmetry that is obtained
from a Drinfeld-Sokolov reduction, for non-critical values of the
central extension, in a generic way by coupling a WZW model
representing the matter fields to a WZW model representing the
(generalized) gravitational degrees of freedom.
The constrained classical models give rise to two separate $\cal W$
algebra realizations, and the constaints entail the presence of
ghosts. A condition for $\cal W$-invariance
of the full theory is always the vanishing of the
sum of the central charges.  We showed (using the field-antifield
formalism) how to derive the BRST charge, always on the classical
level. We showed explicitly the workability of our scheme by applying
it to ${\cal W}_3$. Since the central extensions are already present at the
classical level, the eventual transition to the quantum level was shown
to be rather trivial, involving (in that case) only the renormalization
of a single coefficient, without additional terms. This suggests
that our method may be used to advantage in all these cases where the
transition to the quantum algebra's seems impossible or problematic.
We hope to come back to these questions in the future.

\section*{Acknowledgments}
It is a pleasure to thank Kris Thielemans and Stefan Vandoren
for discussions.

\end{document}